\documentclass[twocolumn,amssymb,prl,showpacs]{revtex4}
\usepackage{epsfig}
\usepackage{graphicx}

\draft
\begin{document}
\title{Proposal for a correlation induced spin-current polarizer}

\author{Piotr Stefa\'nski}
\email{piotrs@ifmpan.poznan.pl}
 \affiliation{Institute of Molecular Physics of the Polish Academy of Sciences\\
  ul. Smoluchowskiego 17, 60-179 Pozna\'n, Poland}
\author{}
\affiliation{}
\author{}
\affiliation{}
\date{\today}

\begin{abstract}
We propose a spin polarizer device composed of a quantum dot
connected to the spin polarized leads. The spin control of the
current flowing through the device is entirely due to the Coulomb
interactions present inside the dot. We show that the initial
polarization present in the source lead can be reverted or
suppressed just by manipulating the gate voltage acting on the
dot, the presence of the external magnetic field is not required.
The influence of the temperature and finite bias on the efficiency
of the current spin switching effect is also discussed.
\end{abstract}
\maketitle

\section{introduction}
The idea of the future electronics based on the spin degree of
freedom instead of charge has emerged during last years. The term
of "spintronics" is one of the most frequently used in the modern
solid state physics \cite{book} in various aspects. Manipulation
of the spin by electric field is an important problem met on the
way to achieve a reasonable alternative to traditional
charge-based semi-conductor electronics. The early work of Datta
and Das suggested the electrical control of the spin utilizing the
Rashba spin-orbit interaction \cite{DattaDas}, which has recently
been realized experimentally \cite{nitta} in a semi-conductor
heterostructure. Spin transport and gate control has also been
realized in carbon nanotubes \cite{saho, jensen}. Recently
half-metallicity has  been induced by external electric field
applied to the graphene nanoribbon \cite{son}.

In the present work we focus on small semiconductor quantum dots
(QD) which offer better scalability and are compatible with
present semiconductor technology. When operated by the gate
voltage in the Coulomb blockade regime, such a QD acts as a
single-electron transistor (SET) \cite{goldhaber}. We will show
that in the presence of ferromagnetic leads SET can invert the
spin of the incoming current due to Coulomb interactions inside
the SET.

To date, many theoretical studies of the interacting dots with
ferromagnetic leads have have been reported
\cite{bulka,krawiec,barnas}. These ideas have been realized
experimentally very recently \cite{hamaya1,hamaya2}.

We show that in the vicinity of degeneracy points at Hubbard
resonances, where the spin-up and spin-down dot occupancies are
equal, the interacting QD in the Coulomb blockade regime can serve
as an effective spin polarizer. We predict two experimentally
promising conditions for spin switching: i) the effect is enhanced
if the dot is asymmetrically coupled to the leads, experimentally
advantageous condition giving a possibility of different switching
fields of the leads \cite{hamaya1,hamaya2}, ii) the current
flowing into the dot should not be fully polarized because the
mechanism of the control of spin polarization is due to the
Coulomb interactions between electrons with opposite spins. Thus,
within the presented proposal, we take advantage of the inevitably
encountered experimental situation, that the resultant current
flowing from spin-polarized electrode into the dot is not 100
percent polarized.

The tunnelling junction between a ferromagnetic metal and 2D
electron gas inside the semiconductor heterostructure
\cite{hamaya1,hamaya2}, which the quantum dot is formed of, can be
approximated by a ferromagnetic-normal metal interface (F/N)
\cite{zutic}. For such a junction the degree of polarization of
the injected current is dependent on  the contact resistance and
the characteristic resistances of F and N components, given by the
ratio of spin diffusion length and effective bulk conductivity of
the corresponding component. Apart from the partial loss of the
spin polarization of the injected current at the junction, there
are several mechanisms of spin relaxation present on the
semiconductor side of the junction \cite{book,zutic}. For confined
structures they originate mainly from the spin-orbit coupling in
the absence of inversion symmetry of the structure and from the
hyperfine interaction between magnetic moments of electrons and
nuclei. In the following we will consider the situation where the
current injected into the dot partially loses its initial
polarization and is not fully polarized even if the source
electrode were. Thus, the polarization of the lead, described
below in terms of $\Gamma_{\sigma}$ widths of the dot level,
should be understood as the effective lead polarization "seen" by
the dot localized state after all spin polarization-loss processes
took place.
\section{Theoretical approach}
The device is described by Anderson hamiltonian \cite{anderson61},
where the dot takes the role of magnetic impurity and the
(polarized) leads are analogues of host metal:
\begin{eqnarray}
\label{for1}  \nonumber H= \epsilon_{d}d_{\sigma}^{+}d_{\sigma}+U
 n_{\sigma}n_{\bar{\sigma}}
 +\sum_{k,\sigma,\alpha=L,R}\lbrack
 t_{\alpha}c_{k\alpha,\sigma}^{+}d_{\sigma}+h.c.]\\+
\sum_{k,\sigma,\alpha=L,R}\epsilon_{k\alpha,\sigma}c_{k\alpha,\sigma}^{+}c_{k\alpha,\sigma}
\end{eqnarray}
The first two terms describe the dot with the presence of Coulomb
interactions $U$. The bare dot level is shifted
 by the gate voltage acting on the dot capacitatively:
$\epsilon_{d}\equiv\epsilon_{d}-V_{g}$, and its initial position
is assumed to coincide with Fermi level
$\epsilon_{d}=\epsilon_{F}=0$. The third term describes the
tunnelling between the dot and the leads, represented by the last
term in Eq.~(\ref{for1}). The electron energy in the leads is
spin-dependent, $\sigma=\uparrow,\downarrow$, because the leads
are assumed to be spin polarized. We neglect the spin dependence
of the tunnelling matrix elements $t_{\alpha}$ ($\alpha=L, R$)
which are rather dependent on the potential barrier between the
dot and a given lead. Thus, the spin dependence of the QD level
width
$(\Gamma_{\sigma}/2)=(1/2)\sum_{\alpha}\Gamma_{\alpha\sigma}$;
$\Gamma_{\alpha\sigma}=2\pi|t_{\alpha}|^{2}\rho_{\alpha\sigma}$ is
caused by the coupling to the leads with different spectral
densities $\rho_{\alpha\uparrow}\neq\rho_{\alpha\downarrow}$,
which are assumed to be featureless and constant.

Let us define the polarization of the quantity $X$,
$P_{X}=(X_{\uparrow}-X_{\downarrow})/(X_{\uparrow}+X_{\downarrow})$.
For the lead $\alpha$ it is:
$P_{\alpha}=(\rho_{\alpha\uparrow}-\rho_{\alpha\downarrow})/(\rho_{\alpha\uparrow}+\rho_{\alpha\downarrow})$,
which can be expressed by the spin-dependent QD widths:
$P_{\alpha}=(\Gamma_{\alpha\uparrow}-\Gamma_{\alpha\downarrow})/(\Gamma_{\alpha\uparrow}+\Gamma_{\alpha\downarrow})$.

The retarded dot Green function
$G_{\sigma}^{r}(t-t')=-i\theta(t-t')\langle
d_{\sigma}(t)d_{\sigma}^{\dagger}(t')+d_{\sigma}^{\dagger}(t')d_{\sigma}(t)\rangle$
is obtained by solving the set of equations of motion of the Green
functions in the Hubbard approximation \cite{hewson}. Within this
approximation the two-particle Green functions describing
spin-flip processes (generating Kondo effect) on the localized
level are neglected. The Green functions that describe the normal
scattering of band electrons on an impurity are approximated by
decoupling of band electrons from impurity electrons. The Hubbard
approximation is valid for large $U/\Gamma$ ratio, when the
Hubbard subbands are well separated in energy scale.  It is the
simplest scheme which describes correlated electrons,  placed on
the approximation scale between Hartree-Fock approximation for
interacting but uncorrelated electrons, and the schemes for
strongly correlated electrons, leading to Kondo physics. Thus, it
is most suitable for the description of a spin-degenerate QD level
in the Coulomb blockade regime of the lead-dot coupling.  The
Fourier-transformed expression for QD Green function with the spin
$\sigma=\uparrow,\downarrow$ has the form:
\begin{eqnarray}
\label{G1gen} \nonumber
G_{\sigma}^{r}(\omega)=[\frac{\omega-\epsilon_{d}}
{1+\frac{\langle n_{\bar{\sigma}}\rangle
U}{\omega-\epsilon_{d}-U}}+\frac{i\Gamma_{\sigma}}{2}]^{-1}\\\simeq
\frac{1-\langle
n_{\bar{\sigma}}\rangle}{\omega-\epsilon_{d}+\frac{i\Gamma_{\sigma}}{2}}+\frac{\langle
n_{\bar{\sigma}}\rangle}{\omega-\epsilon_{d}-U+\frac{i\Gamma_{\sigma}}{2}}.
\end{eqnarray}

Eq. (\ref{G1gen}) has been written as the sum of two Hubbard
resonances, whose spectral weights are controlled by the dot level
occupancy with the opposite spin $\bar{\sigma}$. This feature,
caused by Coulomb interactions between electrons with opposite
spins, is crucial for the spin switching  effect. The occupancies
of spin $\uparrow$ and $\downarrow$ can be very different for the
given gate voltage in spite of degeneracy
$\epsilon_{d\uparrow}=\epsilon_{d\downarrow}$, because of the
different widths of $\epsilon_{d\uparrow}$ and
$\epsilon_{d\downarrow}$  levels introduced by polarized
electrodes. Occupancies have been calculated selfconsistently from
the set of coupled equations:
\begin{eqnarray}
\label{self}
 \nonumber \langle n_{\sigma}\rangle =-\frac{i}{2\pi}\int
G^{<}_{\sigma}(\omega,\langle n_{\bar{\sigma}}\rangle)d\omega,\\
\langle n_{\bar{\sigma}}\rangle =-\frac{i}{2\pi}\int
G^{<}_{\bar{\sigma}}(\omega,\langle n_{\sigma}\rangle)d\omega.
\end{eqnarray}
The "lesser" dot Green function $G^{<}$ can be expressed by the
spectral density of the dot \cite{haug},
$\rho_{\sigma}(\omega)=-(1/\pi)\Im G^{r}_{\sigma}(\omega)$,
$G^{<}_{\sigma}(\omega)=2i\pi\bar{f}(\omega)\rho_{\sigma}(\omega)$.
Non-equilibrium distribution function
$\bar{f}=[\Gamma_{L\sigma}f_{L}+\Gamma_{R\sigma}
f_{R}]/(\Gamma_{L\sigma}+\Gamma_{R\sigma})$ has a two-step profile
defined by the chemical potential in the leads: $f_{L/R}\equiv
f(\omega\mp eV)$ and collapses into equilibrium Fermi-Dirac
distribution function $f\equiv f_{L}=f_{R}$ in the limit of zero
bias between the leads, $eV\rightarrow 0$.
The current is calculated within Landauer formalism from the
relation \cite{haug}:
\begin{equation}
J=\frac{e}{\hbar}\sum_{\sigma}\int d\omega
[f_{L}-f_{R}]\frac{\Gamma_{L\sigma}\Gamma_{R\sigma}}{\Gamma_{L\sigma}+\Gamma_{R\sigma}}\rho_{\sigma}(\omega).
\end{equation}
In the limit of zero bias the conductance has the form:
\begin{equation}
G=\frac{\partial J}{\partial V}
=\frac{2e^2}{\hbar}\sum_{\sigma}\int d\omega (-\frac{\partial
f}{\partial
\omega})\frac{\Gamma_{L\sigma}\Gamma_{R\sigma}}{\Gamma_{L\sigma}+\Gamma_{R\sigma}}\rho_{\sigma}(\omega).
\end{equation}
\section{Results and discussion}

\begin{figure} [ht]
\epsfxsize=0.45\textwidth \epsfbox{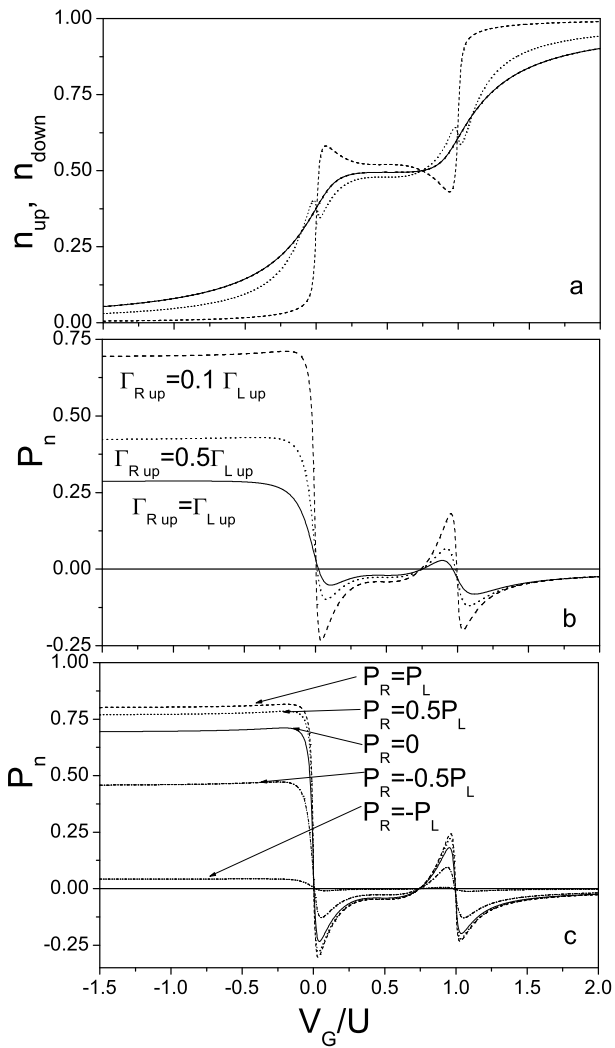}
\caption{\label{fig1}Panel (a): occupancies $\langle n_{\uparrow}
\rangle$ and $\langle n_{\downarrow}\rangle$ vs. gate voltage for
zero temperature, $P_{L}=0.8$ and $P_{R}=0$ and asymmetric
coupling to the leads $\Gamma_{R\uparrow}=0.1\Gamma_{L\uparrow}$
(dotted curve-$n_{\uparrow}$, dashed curve- $n_{\downarrow}$). The
solid curve shows the occupancy $n_{\uparrow}=n_{\downarrow}$ for
unpolarized leads. Panel (b) shows the polarization of the dot for
the same leads polarization as in (a), but with the decrease of
the asymmetry of the coupling to the leads. Panel (c) shows the
polarization of the dot for the same parameters as for (a), but
when the right lead polarization changes. }
\end{figure}

When the leads are unpolarized, $P_{L}=P_{R}=0$, the occupancy
curves for $n_{\uparrow}$ and  $n_{\downarrow}$ coincide and the
usual plateau of the width $\sim U$ appears when the first
$\epsilon_{d}$ and second $\epsilon_{d}+U$ Hubbard levels are
filled with electrons when gate voltage changes (see solid curve
in Fig.~(\ref{fig1}a)). An
 introduction of the spin-polarized leads changes the situation.
The non-monotonicity of the dot spin-up and spin-down occupations
appears with the increase of the left electrode polarization (we
focus on the case when right lead polarization $P_{R}=0$ and
asymmetric QD-leads coupling
$\Gamma_{R\uparrow}=0.1\Gamma_{L\uparrow}$ is assumed in present
discussion, unless stated differently \cite{furtherwork}). Now the
spin-dependent widths of $\epsilon_{d}$ level
$\Gamma_{\uparrow}\neq \Gamma_{\downarrow}$, which introduces the
difference in the $n_{\uparrow}$ and $n_{\downarrow}$ as
calculated from the integration of the corresponding spectral
densities (see Eq.~(\ref{self})). The weights of the spectral
peaks of $\rho_{\uparrow}$ and $\rho_{\downarrow}$  become
different as controlled by the occupancy of opposite spins,
Eq.~(\ref{G1gen}).

The present model is formally equivalent to the model of spinless
electron double-dot system with Coulomb interaction between the
dots \cite{koenig} for the case of dots levels degeneracy and
anisotropy of the levels coupling to the leads. Within this model
non-monotonicity in the occupancy has also been encountered.

There are three degeneracy points where
$n_{\uparrow}=n_{\downarrow}$ shown in Fig.~(\ref{fig1}a). Two of
them are for the gate voltages when the levels $\epsilon_{d}$
($V_{g}=0$) and $\epsilon_{d}+U$ ($V_{g}=U$) respectively coincide
with Fermi energy. They are the most advantageous for the spin
control.  The third point, when $\langle n_{\uparrow} \rangle
=\langle n_{\downarrow}\rangle\cong 0.5$ appears when the Fermi
level is placed between the Hubbard levels. For unpolarized
electrodes it corresponds to the symmetric Anderson model when
$\epsilon_{d}=-U/2$ ($V_{g}=U/2$) \cite{anderson61}.

For the remaining gate voltages  $\langle n_{\uparrow}\rangle\neq
\langle n_{\downarrow}\rangle$. In the limiting case of $P_{L}=1$
the occupancy curves do not change much, the spin down electrons
are still supplied to the dot from the right, unpolarized lead to
modify the spectral density $\rho_{\uparrow}$ (see
Eq.~(\ref{G1gen})) and $\langle n_{\uparrow}\rangle$ occupancy as
a result of Coulomb interactions.

The Panel (b) of Fig. (\ref{fig1}) shows the change of the dot
occupancy polarization $P_{n}$ with decrease of the asymmetry of
the coupling to the leads. The efficiency of the switching of the
initial polarization decreases for more symmetric dot-leads
coupling, but the device is still operational even for symmetric
coupling.

The Panel (c) shows $P_{n}$ for various right lead polarizations
$P_{R}$. $P_{n}$ is robust in the range of $V_{g}$ where Coulomb
interaction inside the dot plays the role in the spin switching
until the right lead has an opposite polarization with respect to
the left one. In such a case the effect is strongly diminished
\cite{furtherwork}.

\begin{figure} [ht]
\epsfxsize=0.45\textwidth \epsfbox{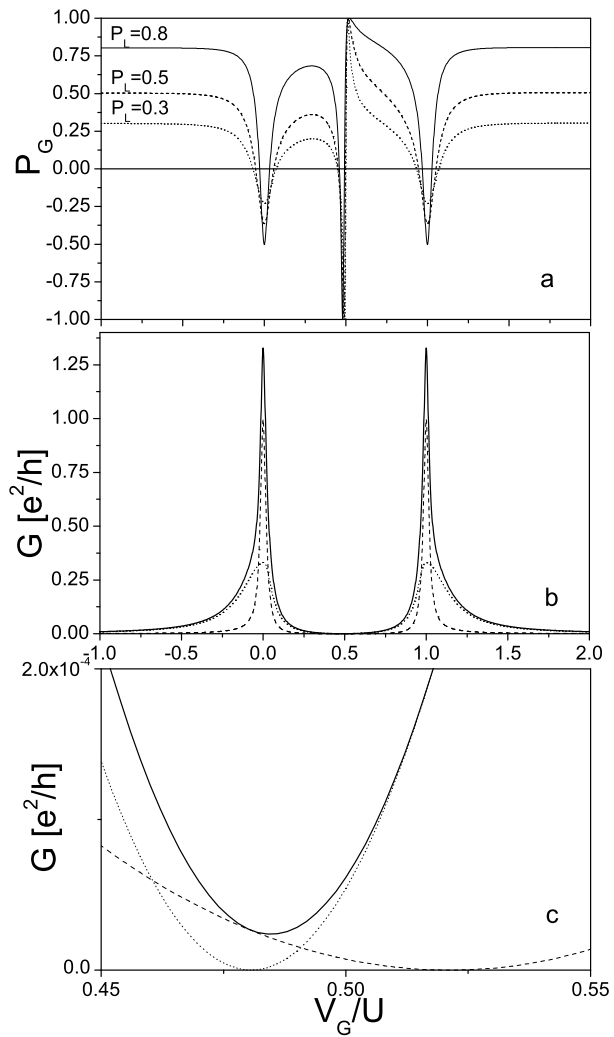}
\caption{\label{fig2}Panel (a): zero-bias conductance polarization
calculated for the same parameters as for Fig.~(\ref{fig1}a)  but
for various initial source lead polarizations. Panel (b) shows the
conductance for $P_{L}=0.8$ with its spin-dependent components:
dashed line- spin down and dotted line- spin up components,
respectively. Bold solid line is the total conductance. Panel (c):
magnification of the region in-between conductance peaks showed in
(b), note the change of the gate voltage scale.}
\end{figure}

Zero bias conductance polarization, $P_{G}$, is shown in
Fig.~(\ref{fig2}a) for various $P_{L}$ values. There are three
gate voltage ranges for which the polarization of the initial
current flowing from the source can be reverted. The most
favorable for operation are the values $V_{g}=0$ and $V_{g}=U$,
which correspond to high values of conductance at the Coulomb
peaks (see Panel (b)). The third point at $V_{g}=U/2$ is less
favorable because it is placed in-between Coulomb blockade peaks,
where the conductance is very small. In this point, regardless of
the initial $P_{L}$, $P_{G}$ reaches the value of -1 and then
switches to the value of +1. This feature can be understood when
analyzing the expression for zero-bias conductance. At $T=0$, its
$\sigma$-component has the form:
\begin{equation}
G_{\sigma}=\frac{2e^2}{h}\frac{\Gamma_{L\sigma}\Gamma_{R\sigma}}{(\frac{\epsilon_{d}(\epsilon_{d}+U)}{\epsilon_{d}+U(1-\langle
n_{\bar{\sigma}}\rangle)})^{2}+(\Gamma_{\sigma}/2)^2}.
\end{equation}
It reaches zero value when the denominator
$\epsilon_{d}+U(1-\langle n_{\bar{\sigma}}\rangle)=0$, which for
unpolarized leads gives symmetric case: $\epsilon_{d}=-U/2$ and
$\langle n_{\bar{\sigma}}\rangle=\langle n_{\sigma}\rangle=0.5$.
For polarized leads the position of $\epsilon_{d}$ giving
$G_{\sigma}=0$ is different for each conductance spin component
because $\langle n_{{\sigma}}\rangle\neq \langle
n_{\bar{\sigma}}\rangle$. In our case, for $P_{L}>0$,
$n_{\downarrow}>n_{\uparrow}$ (Fig.~(\ref{fig1}a)) in-between CB
peaks and $G_{\uparrow}=0$ for smaller value of $V_{g}$ than
$G_{\downarrow}=0$, which implies $P_{G}=-1$ for such gate voltage
(see Fig.~(\ref{fig2}c)). When the gate voltage increases further,
$G_{\downarrow}$ in turn reaches zero value and the conductance
polarization jumps to the value of +1. The values of
$n_{\downarrow}$ and $n_{\uparrow}$  in the region between CB
peaks are weakly dependent on the initial $P_{L}$ polarization and
the difference between them is small. It causes the same weak
dependence on $P_{L}$ of the  of zero-bias conductance
polarization in this region.

Panel (b)  of Fig.~(\ref{fig2}) shows the conductance with its
spin-dependent contributions for $P_{L}=0.8$ and
$\Gamma_{R\uparrow}=0.1\Gamma_{R\downarrow}$.  The total
conductance does not reach unitary limit because of the asymmetry
of the coupling of the dot to the leads. The conductance peaks are
very asymmetric due to the peculiar behavior of the occupancies
and coupling asymmetry.

When the sign of $P_{L}$ is changed, the obtained polarization
curves are reverted with respect to conductance polarization
$P_{G}=0$ line (not shown).

The effect of conductance polarization switching is more
pronounced when the dot is asymmetrically coupled to the leads,
promoting better control of the leads polarizations by different
switching fields \cite{hamaya1, hamaya2}. Moreover, the asymmetric
coupling is naturally accessible experimentally, contrary to the
fully symmetric coupling which needs some tuning of the quantum
point contacts between the leads and the dot.

The current polarization change at the resonances for $V_{g}=0$
and for $V_{g}=U$ offers a new electron correlations-based
mechanism for the change of the sign of tunnelling
magnetoresistance by the gate voltage, which has recently been
observed \cite{hamaya2}.

The polarization $P_{G}$ dependence vs. gate voltage does not
exactly match the quantum dot polarization, especially in the gate
voltage range where the dot is occupied by one electron. This is
due to the fact that the conductance polarization is not directly
related to the dot polarization but rather to the value of the
spin-dependent QD spectral densities at the Fermi level.

In the limiting case of $P_{L}=1$ the conductance polarization is
always $P_G=1$ (contrary to the dot occupancy polarization
discussed above) because $\Gamma_{L\downarrow}=0$ and the
spin-down contribution to the conductance is switched off. The
effectiveness of current polarization switching by the Coulomb
blockaded SET is operative in realistic situations, when the
electrons incoming to the dot are not 100 percent polarized.

The current polarization switching is robust to the temperature
increase in the regions of the operation of the device- for
$V_{g}\sim 0$ and $V_{g}\sim U$, as compared to the point of
$V_{g}\sim U/2$.
\begin{figure} [ht]
\epsfxsize=0.45\textwidth \epsfbox{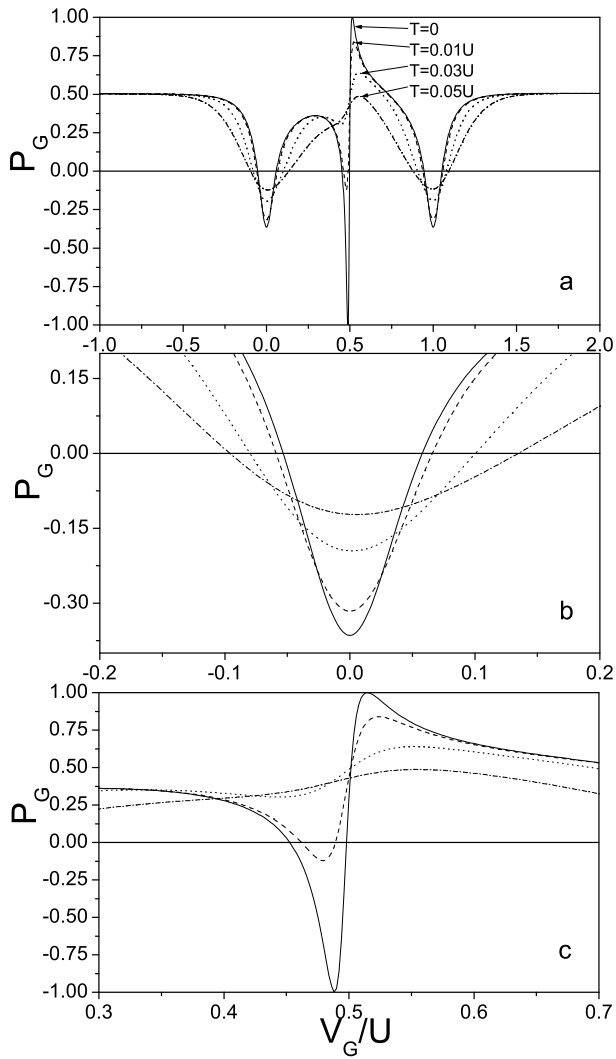}
\caption{\label{fig3}Panel (a): zero-bias conductance polarization
calculated for various temperatures, $P_{L}=0.5$, $P_{R}=0$ and
$\Gamma_{L\uparrow}=0.1\Gamma_{R\uparrow}$. Panels (b) and (c)
show the magnification of regions in the vicinity of $V_{g}=0$ and
$V_{g}=0.5U$, respectively. Note the changes of the gate voltage
scale.}
\end{figure}
\begin{figure} [ht]
\epsfxsize=0.45\textwidth \epsfbox{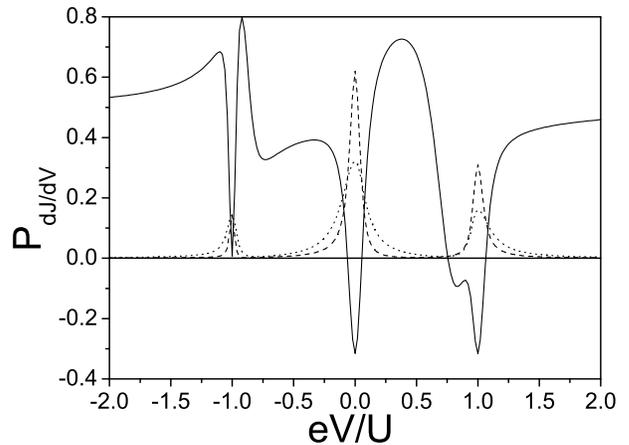}
\caption{\label{fig4}Bias dependence of the differential
conductance polarization (solid line) calculated for $V_{g}=0$ and
temperature $T=0.01U$, $P_{L}=0.5$, $P_{R}=0$ and
$\Gamma_{L\uparrow}=0.1\Gamma_{R\uparrow}$. The spin-dependent
contributions to the conductance are shown: spin down- dashed line
and spin up- dotted line. }
\end{figure}
It is shown in Fig.~(\ref{fig3}) for the source lead polarization
$P_{L}=0.5$. For the temperature $T=0.01U$ ( $=174 mK$ for $U=15
meV$ \cite{hamaya2}), which is a typical range for the SET
operation in the Coulomb blockade
\cite{goldhaber,hamaya1,hamaya2}, $P_{G}\simeq -0.35$ at $V_{g}=0$
shown in the Panel (b) of Fig. (\ref{fig3}). For $T=0.05U$ the
switching is less efficient, but still present. The effect is
limited rather by the temperature value for which the Coulomb
blockade becomes visible $k_{B}T < \Gamma < U$. Contrary, the
$P_{G}$ is sensitive to the temperature change in the region close
to $V_{g}=0.5U$,  Panel (c) of Fig.~(\ref{fig3}). The increase of
thermal broadening of the conductance peaks causes the decrease of
the difference between spin up and spin down components in this
region, which is reflected in rapid decrease of polarization.

The dependence of the differential conductance polarization
$P_{dJ/dV}$ on the applied bias voltage calculated at $V_{g}=0$ is
shown in Fig.~(\ref{fig4}) along with conductance spin-dependent
components for the temperature $T=0.01U$ and $P_{L}=0.5$. At
$V_{g}=0$ the dot $\epsilon_{d}$ level matches the effective
chemical potential in the leads for zero bias. The switching
effect decreases for finite bias and approaches $P_{dJ/dV}=0$ for
$eV \sim \pm 0.05U$. There are additional $P_{dJ/dV}$ polarization
anomalies, which appear for larger values of applied bias, $eV=\pm
U$. Namely, for $eV=-U$ the chemical potential in the left lead
comes into resonance with the $\epsilon_{d}+U$ level, which is
reflected by the maxima shown in Fig.~(\ref{fig4}) of spin
components of conductance for such a bias. Similarly, for $eV=U$
the chemical potential of the right lead is in resonance with
$\epsilon_{d}+U$ level. The large bias spin transport has indeed
been observed very recently \cite{hamaya1}. The anomalies at large
bias are sensitive to the leads polarization arrangement and
asymmetry of the dot-leads coupling \cite{furtherwork}.

In conclusion, we have shown that the spin polarization of the
current can be inverted electrically by the gate voltage acting on
the SET in Coulomb blockade regime. The effect is purely due to
Coulomb interactions present inside the QD. Current polarization
switching is robust to the temperature change and favored by
inevitably encountered experimental conditions: asymmetry of the
dot-lead coupling and partial lose of the initial current
polarization at the dot-lead interface. The model sheds also new
light on layered spin-polarized heterostructures \cite{ohno}
operated by external electric field, where the bound states can be
formed inside the interface due to the spatial confinement and
energy structure mismatch of heterostructure components.
\begin{acknowledgements}
I acknowledge valuable suggestions from A. Tagliacozzo. The work
is supported from the European Science Foundation EUROCORES
Programme FoNE by funds from the Ministry of Science and Higher
Education and EC 6FP (contract N. ERAS-CT-2003-980409) and the EC
project RTNNANO (contract N. MRTN-CT-2003-504574).
\end{acknowledgements}


\begin{thebibliography}{99}
\bibitem{furtherwork} The detailed discussion of the device behavior for finite
applied bias and various polarizations of the electrodes will be
presented in the subsequent publication.
\end{thebibliography}
\end{document}